\documentclass{emulateapj}
\usepackage{natbib}
\usepackage{graphicx}
\usepackage{amsmath}
\usepackage{color}

\shorttitle{CLASP-IRIS}
\shortauthors{Schmit et al.}

\newcommand{\HI}{\ion{H}{1}}
\newcommand{\MgII}{\ion{Mg}{2}}

\begin{document}

\title{Comparison of Solar Fine Structure Observed Simultaneously in Ly-$\alpha$ and Mg II h}
\author{D. Schmit\altaffilmark{1,2}}
\author{A. V. Sukhorukov \altaffilmark{3,4}}
\author{B. De Pontieu \altaffilmark{2,5}}
\author{J. Leenaarts \altaffilmark{3}}
\author{C. Bethge \altaffilmark{6,7}}
\author{A. Winebarger \altaffilmark{6}}
\author{F. Auch\`ere \altaffilmark{8}}
\author{T. Bando\altaffilmark{9}}
\author{R. Ishikawa\altaffilmark{10}}
\author{R. Kano\altaffilmark{9}}
\author{K. Kobayashi\altaffilmark{6}}
\author{N. Narukage\altaffilmark{9}}
\author{J. Trujillo Bueno\altaffilmark{11}}

\affil{Bay Area Environmental Research Institute, 625 2nd St. Ste 209 Petaluma, CA 94952, USA$^1$}

\affil{Lockheed Martin Solar and Astrophysics Laboratory, Building 252, 3176 Porter Dr, Palo Alto, CA 94304, USA$^2$}

\affil{Institute for Theoretical Astrophysics, University of Oslo, PO Box 1029, Blindern 0315 Oslo Norway$^3$}

\affil{Institute for Solar Physics, Dept. of Astronomy, Stockholm University, SE-106 91 Stockholm, Sweden$^4$}

\affil{Main Astronomical Observatory,
National Academy of Sciences of Ukraine,
27 Akademika Zabolotnoho str., 03680 Kyiv, Ukraine$^5$}

\affil{NASA Marshall Space Flight Center, ZP 13, Huntsville, AL 35812, USA$^6$}

\affil{Universities Space Research Association, 320 Sparkman
Drive, Huntsville, AL, 35806, USA$^7$}

\affil{Institut d'Astrophysique Spatiale, CNRS/Univ. Paris-Sud 11, B\^atiment 121, F-91405 Orsay, France$^8$}

\affil{National Astronomical Observatory of Japan, National Institutes of Natural Sciences, 2-21-1 Osawa, Mitaka, Tokyo 181-8588, Japan $^9$}

\affil{Institute of Space and Astronautical Science, Japan Aerospace Exploration Agency, 3-1-1 Yoshinodai, Chuo-ku, Sagamihara, Kanagawa 252-5210, Japan $^10$}

\affil{Instituto de Astrof\'{i}sica de Canarias, E-38205 La Laguna, Tenerife, Spain$^{11}$}

\begin{abstract}
The Chromospheric Lyman Alpha Spectropolarimeter (CLASP) observed the Sun in H I Lyman-$\alpha$ during a suborbital rocket flight on September 3, 2015.
The Interface Region Imaging Telescope (IRIS) coordinated with the CLASP observations and recorded nearly simultaneous and co-spatial observations in the Mg II h\&k lines.
The Mg II h and Ly-$\alpha$ lines are important transitions, energetically and diagnostically, in the chromosphere.
The canonical solar atmosphere model predicts that these lines form in close proximity to each other and so we expect that the line profiles will exhibit similar variability.
In this analysis, we present these coordinated observations and discuss how the two profiles compare over a region of quiet sun at viewing angles that approach the limb.
In addition to the observations, we synthesize both line profiles using a 3D radiation-MHD simulation.
In the observations, we find that the peak width and the peak intensities are well correlated between the lines.
For the simulation, we do not find the same relationship.
We have attempted to mitigate the instrumental differences between IRIS and CLASP and to reproduce the instrumental factors in the synthetic profiles.
The model indicates that formation heights of the lines differ in a somewhat regular fashion related to magnetic geometry.
This variation explains to some degree the lack of correlation, observed and synthesized, between Mg II and Ly-$\alpha$.
Our analysis will aid in the definition of future observatories that aim to link dynamics in the chromosphere and transition region.
\end{abstract}

\section{Introduction}
It has been known since the successful identification of the Fe XIV green line \citep{edlen_43} that the solar atmosphere undergoes a transition in the upper atmosphere, where an unresolved heat source produces a hot corona.
The 1 MK corona is connected to the photosphere through a dynamic and relatively-minute vertical swath of atmosphere, namely the transition region and chromosphere.
Radiation plays a fundamental role in the energy balance here.
Many of the important atomic transitions from this region occur in the UV and are not accessible to ground-based remote sensing instruments.\\
%The space age has been a boon for our understanding.
\indent The Interface Region Imaging Spectrograph satellite (IRIS) has provided us a wealth of new information derived from the Mg II h\&k lines at 2803.5\AA~and 2796.4\AA~respectively \citep{depontieu_14}.
At the base of the transition region, the most diagnostically important is undoubtedly H I Lyman-$\alpha$ at 1215.7\AA.
Despite its importance, there are preciously few measurements of Ly-$\alpha$.
The line is bright and broad.
The most recent spectroscopic measurements were made with SUMER \citep{curdt_08}, but these observations were never routinely made due to the concern of damaging of the detector.
The SUMER dataset is valuable in that include Ly-$\beta$ profiles as well.
Prior to SUMER, the OSO-8 spacecraft was host to the LPSP spectrograph which observed both Mg II h\&k \citep{bocch_94} and Ly-$\alpha \beta$ \citep{lemaire_78} between 1975 and 1978.
The interpretation of Ly-$\alpha$ relies on radiative transfer models, and these models are an active research topic in solar physics \citep{hubeny_95, avrett_08}.\\
\indent In the datasets presented here, we have one of the few coordinated datasets with both Mg II h\&k and Ly-$\alpha$ spectral profiles.
These data provide a unique probe into the connection between the transition region and the chromosphere. 
In order to understand the relationship between properties of the emergent profiles and the emitting plasma, a model atmosphere and detailed radiative transfer calculations are required.
We use the Bifrost code \citep{gudiksen_11} to produce a model atmosphere and compute synthetic line profiles with the Multi3d code \citep{leenaarts_09b,sukhorukov_17}
Thus, we compare not only our two observed lines, but the two synthesized lines as well to ascertain the lines are linked in the atmosphere.
In Section 2, we will describe the instruments, the observing program, the co-alignment of the datasets, and the properties of the Bifrost MHD model.
In Section 3, we analyze the positional variations and statistical connections between the two lines.
In Section 4, we summarize our results.
\begin{figure*}
\centering
\includegraphics[width=0.8\textwidth,trim=3cm 11.5cm 3cm 2cm,clip]{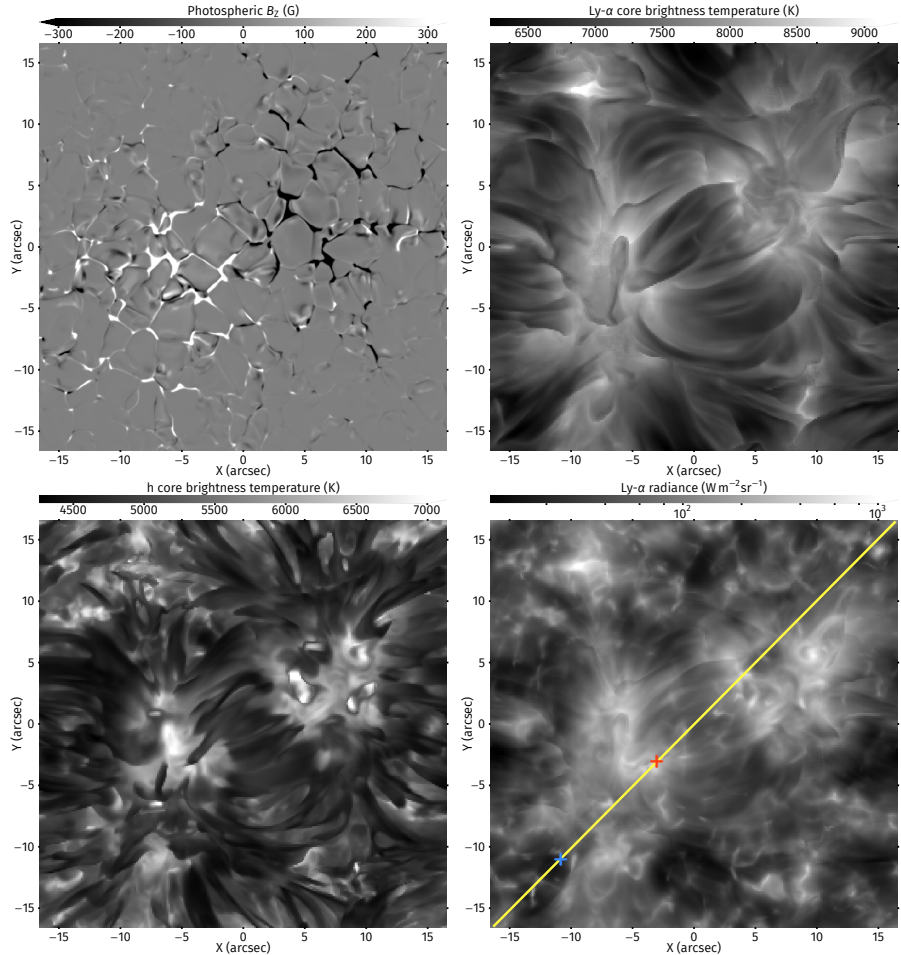}
\caption{Illustration of the Bifrost snapshot and the vertically emergent intensity in the \HI\ Ly-$\alpha$ and \MgII~h lines. Upper left: vertical magnetic field strength in the photosphere. Upper right: vertically emergent intensity at the nominal line core of Ly-$\alpha$. Lower left: vertically emergent intensity at the nominal line core of  \MgII~h. Lower right: vertically emergent frequency-integrated Ly-$\alpha$ intensity. The yellow line shows the slice extracted and displayed in Figure 9. The red and blue crosses show the location of the network and internetwork profiles displayed in Figure 4.}
\label{fig:sim}
\end{figure*}
\section{Observations}
\subsection{CLASP}
\begin{figure}
\includegraphics[width=0.5\textwidth]{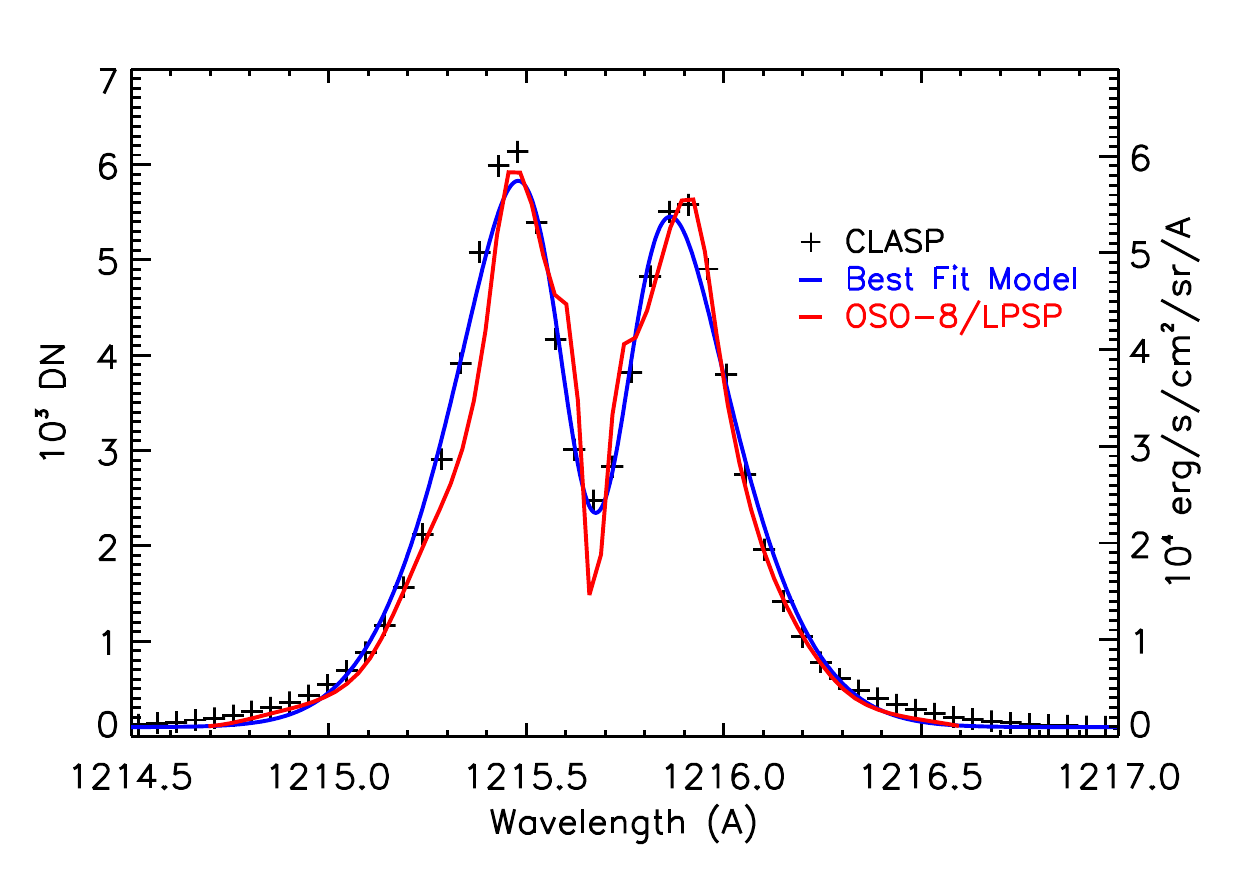}
\caption{Comparison of Ly-$\alpha$ profiles. CLASP data is the average of 20" of profiles at $\mu$=0.75. The best fit model was determined using the method discussed in Section 2.1. The OSO-8 profile was extracted from \cite{gout_78}.}
\label{fig:ex_ly}
\end{figure}

The Chromospheric Lyman Alpha Spectropolarimter (CLASP) is a rocket-borne spectropolarimeter and slit jaw camera \citep{kano_12,kobayashi_12}.
The optical systems have been designed for high throughput in the ultraviolet to observe the H I Ly-$\alpha$ line.
The primary science objective of the mission is to measure the radiation polarization fraction across the profile as a diagnostic of the upper-chromospheric magnetic field \citep{trujillo_11}.
With that objective in mind, high spectral resolution is less desirable than good photon statistics.\\
\indent The CLASP rocket collected data for approximately 322 seconds of its flight September 3, 2015 \citep{kano_17}.
Two pointings were conducted during that interval.
For the first 45 seconds, the instrument pointed disk center to collect calibration data.
The instrument then slewed to a quiet sun target near the limb with the slit oriented radially with respect to disk center so that a maximum breadth of viewing angles ($\mu=\cos\theta$) were observed.
The quiet sun target was acquired at 17:03:41 UT.
IRIS coordinated its observing program to overlap with the second pointing, and it is data from this period that will be discussed here.\\
\indent There are two complications that need to be considered in interpreting the CLASP spectral data.
First, geocoronal absorption occurs in the core of the profile.
This is due to hydrogen atoms that occur at low abundance throughout the Earth's upper atmosphere.
Aeronomy measurements provide a baseline value of the column density as 10$^{14}$ cm$^{-2}$ at an altitude of 200 km with temperatures of order 10$^3$ K \citep{banks_73}.
%The thermal width of this absorption feature is less than the spectral plate scale of CLASP.
We have not attempted to remove the geocoronal component because the spectrograph does not resolve it and the width and depth of the line are only weakly constrained.\\
\indent There is a second absorption effect present in the data due to an operational anomaly.  During data collection, water vapor was trapped in the spectrograph section of the instrument. Ice that
accumulated on camera cooling lines prior to launch sublimated after launch.
A similar absorption was reported by \cite{blamont_69}.
This vapor was trapped in the spectrograph due to an inadequate vent between the spectrograph and telescope sections. H$_2$O can be dissociated by Ly-$\alpha$ photons \citep{lewis_83}, and this creates a wavelength-dependent extinction in the observed profiles.  The density of the water vapor increased during the flight, which caused the extension to be time-dependent as well.   Given the complex interaction between radiation and all the molecular permutations of oxygen and hydrogen, we do not attempt to correct for this absorption in this paper. Instead, we present the data here as measured. We are cautious in our interpretation of the data in Section 3 and describe the limitations of the data.  See Winebarger et al. (2017, in prep) for more detailed discussion of the water vapor absorption in the CLASP data set.
%H$_2$O can be dissociated by Ly-$\alpha$ photons \citep{lewis_83}, and this would create a time-dependent extinction in the observed profiles.
%Our Ly-$\alpha$ profiles are approximately symmetric and match previous observations well \citep{bonnet_78,gout_78}, implying that the absorption is smooth over the width of the Ly-$\alpha$ profile.
\\
%Our dataset is unique however and for this reason it is important to gleam from it the information that we can.\\
\indent The CLASP instrument observes the Sun using both a slit jaw camera and a spectrograph.
The spectrograph is described in \cite{ishikawa_14}.
%The slit jaw filter and camera are described in cite[].
The slit jaw camera took 0.6~s exposures at 0.6~s cadence.
The slit jaw camera uses a spectral bandpass filter centered on 1215\AA~with a FWHM transmission of 35\AA~\citep{kubo_16}.
The spectrograph camera took 0.3~s exposures at 0.3~s cadence.
The slit position on the Sun is held approximately constant with a 1" drift over the duration of the observation.
A half wave plate, mounted between the secondary mirror and the slit prism, rotates at a precise and constant rate of 1.3 rad s$^{-1}$.
By coadding 4 sequential frames, we recover a signal that is pure Stokes I.
Two cameras record the $m=\pm$1 order spectra.
We have conducted our analysis using the Camera 1 data.
The optical specifications for CLASP are described in \cite{giono_16}.
The CLASP slit is 1.44" wide.
The spectral resolution of the spectrograph is estimated at 112 m\AA~(FWHM based on pre-flight measurements) with 48 m\AA~platescale.
The spatial resolution of the spectrograph is estimated to be 2.8" with a 1.1" platescale.
The slit jaw data has a spatial resolution of 2.1" with a 1.03" plate scale.
Due to the anomalous absorption, we do not have an absolute calibration of the spectrograph intensities.
We have normalized the data to match the radiance observed by LPSP (disk center mean profile from \citealt{gout_78}).\\
\indent We have chosen to parameterize the Ly-$\alpha$ dataset using a profile fitting technique to facilitate statistical analysis.
This technique has been previously applied to Mg II profiles in \cite{schmit_15}.
For each Ly-$\alpha$ profile, a best-fit model for the parameters $a, b, c, d, f, g$ and $h$ of the form:
\begin{equation}
\begin{aligned}
I(\lambda)=a+b \bigg[ \exp \left( \frac{-(\lambda-c)^2}{d^2}\right)\\
-f \exp \left( \frac{-(\lambda-g)^2}{h^2}\right) \bigg]
\end{aligned}
\end{equation}
is found through a least squares Levenberg-Marquardt minimization routine, MPFIT \citep{markwardt_09}.
This model is capable of producing both asymmetric single-peaked profiles or double-peaked profiles with a depressed core,  depending on the parameter values.
Figure \ref{fig:ex_ly} shows an example of an averaged profile and best-fit model.
The intensity extrema for Mg II h are referenced in the following scheme: h2v (violet peak), h3 (core), and h2r (red peak).
We reference the analogous features on the Ly-$\alpha$ profile similarly: L2v (violet peak), L3 (core), L2r (red peak).
For Ly-$\alpha$, the geocoronal absorption profile does affect the fit.
While the derived L3 wavelength is variable, the magnitude of its variability is reduced via the convolution with the geocoronal profile.\\ 
\indent The peak width is the spectral separation of the v- and r-peaks (sometimes referred to as peak-to-peak distance).
The peak asymmetry is defined as:
\begin{equation}
R_{Ly}=\frac{I(\lambda_\mathrm{L2v})-I(\lambda_\mathrm{L2r})}{I(\lambda_\mathrm{L2v})+I(\lambda_\mathrm{L2r})}.
\end{equation}
which is completely congruent with the Mg II statistic in \cite{schmit_15}.
%While the typical Mg II h profile has a local intensity minima at $\pm0.6$\AA, no such minima exist for Ly-$\alpha$.
%For Mg II, base width was define as the spectral separation of h1v and h1r.
%For Ly-$\alpha$, we define the base width as the separation of the points where the model profile intensities match the following condition:
%\begin{equation}
%I(\lambda_{L1})=a+0.05*(I(\lambda_{h2v})+I(\lambda_{h2r})-2a).
%\end{equation}
%The core depth and asymmetry statistics are defined identically for Mg II and Ly-$\alpha$.
%Depth is defined as:
%\begin{equation}
%D_{Ly}=1-\frac{2I(\lambda_{L3})}{I(\lambda_{L2v})+I(\lambda_{L2r})}.
%\end{equation}
\begin{figure*}
\centering
\includegraphics[width=.9\textwidth]{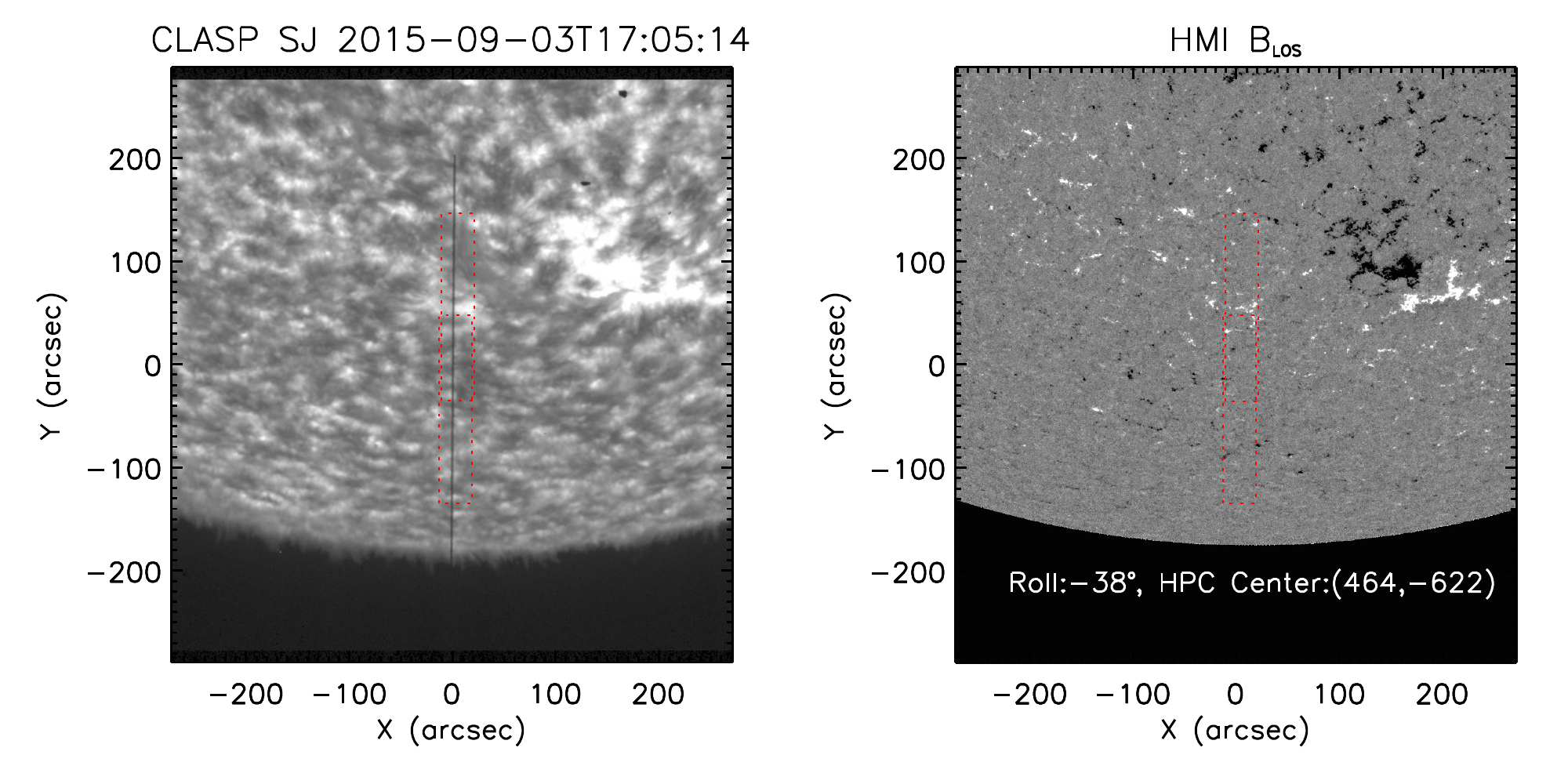}
\caption{CLASP Ly-$\alpha$ slit jaw (left) and the aligned HMI line of sight magnetogram (right). The vertical black line in the left panel is the CLASP slit. The IRIS raster bounds are shown in red.  }
\label{fig:fov}
\end{figure*}
\subsection{IRIS}
The IRIS instrument is described in detail in \cite{depontieu_14}.
IRIS rolled 37$^\circ$ relative to Solar North to position the slit parallel to the CLASP slit.
In the discussion below, the X and Y vectors are defined perpendicular and parallel to the slit respectively.
The IRIS spectrograph (SG) cameras read out a 175" long region of CCDs in the Y-direction.
Spatial binning produced an effective pixel size of 0.33" in the Y-direction.
The IRIS spectrograph data has been converted into physical units based on the Solarsoft routine IRIS\_GET\_RESPONSE (version 3).
The IRIS slit is 0.33" wide.
The IRIS spectrograph has 53 m\AA~resolution with 25 m\AA~platescale.
In order to maximize overlap with the CLASP slit, IRIS conducted a four-stage rastering program.
In stage 1, IRIS takes 1s exposures and makes +1" steps in the X-direction for 32 steps.
In stage 2, IRIS repositions the slit +100" in the Y-direction.
Stage 3 is identical to stage 1, but at the new Y-position.
In stage 4, IRIS repositions the slit  -100" in the Y-direction.
This cycle was repeated 50 times (beginning at 16:29:24 UT) and takes 147 s to complete.
The IRIS slit jaw (SJ) cameras recorded one exposure for every two steps of the X-direction raster.
Only the 1400\AA~bandpass filter was used.
\subsection{Alignment}
The basic alignment of the IRIS and CLASP datasets was done using the SJ data from both instruments.
Ly-$\alpha$ and Si IV 1393\AA~and 1402\AA~are chromospheric/lower transition region spectral lines and the SJ data appear similar between instruments near magnetic concentrations.
We estimate the accuracy of the SJ alignment at 1".
There is a magnification difference between the CLASP SJ and spectrograph data, but the finite extent of the slit is used to map the spectrograph row coordinates to Y-position.
%The CLASP SG spatial plate scale is 1.1".
The IRIS SG data was aligned using spectroheliograms in Si IV 1393\AA, Mg II h 2803\AA, and the SJ 1400\AA~data.
The FUV spectrum signal is only detectable near magnetic concentrations.
We estimate the accuracy of the IRIS SG and SJ alignment at 0.33".
%%%%%%%%%%%%%%
\subsection{Bifrost Model}
%%%%%%%%%%%%%%

%%%%%%%%%%%
%%%%%%%%%%%%%%%%
To better understand our observed spectra, we calculated synthetic spectra of the \HI\  and the \MgII\ atoms from a 3D radiation-MHD simulation, computed using the Bifrost code 
\citep{gudiksen_11}.
Bifrost solves the equations of resistive MHD, together with non-LTE radiative losses in the photosphere and chromosphere, optically thin losses in the corona, and heat conduction along field lines. We used a snapshot from the ``cb24bihe-halfxy-100" run, which was also used in
\citet{leenaarts_16}
and
\citet{golding_17}.
The simulation box spans from the upper convection zone up to the lower corona, with an horizontal extent of $24\times24$~Mm and a vertical extent of 16.8~Mm, from 2.5~Mm below the photosphere to 14.3~Mm above it. The simulation had a grid size of $504 \times 504 \times 496$. Bifrost can run with different equations of state (EOS). The run that we use was run with an EOS that takes non-equilibrium ionization fo hydrogen and helium into account, which leads to a more realistic temperature and electron density structure in the chromosphere and transition region.
This EOS is described in detail in
\begin{figure}
\centering
\includegraphics[width=0.42\textwidth]{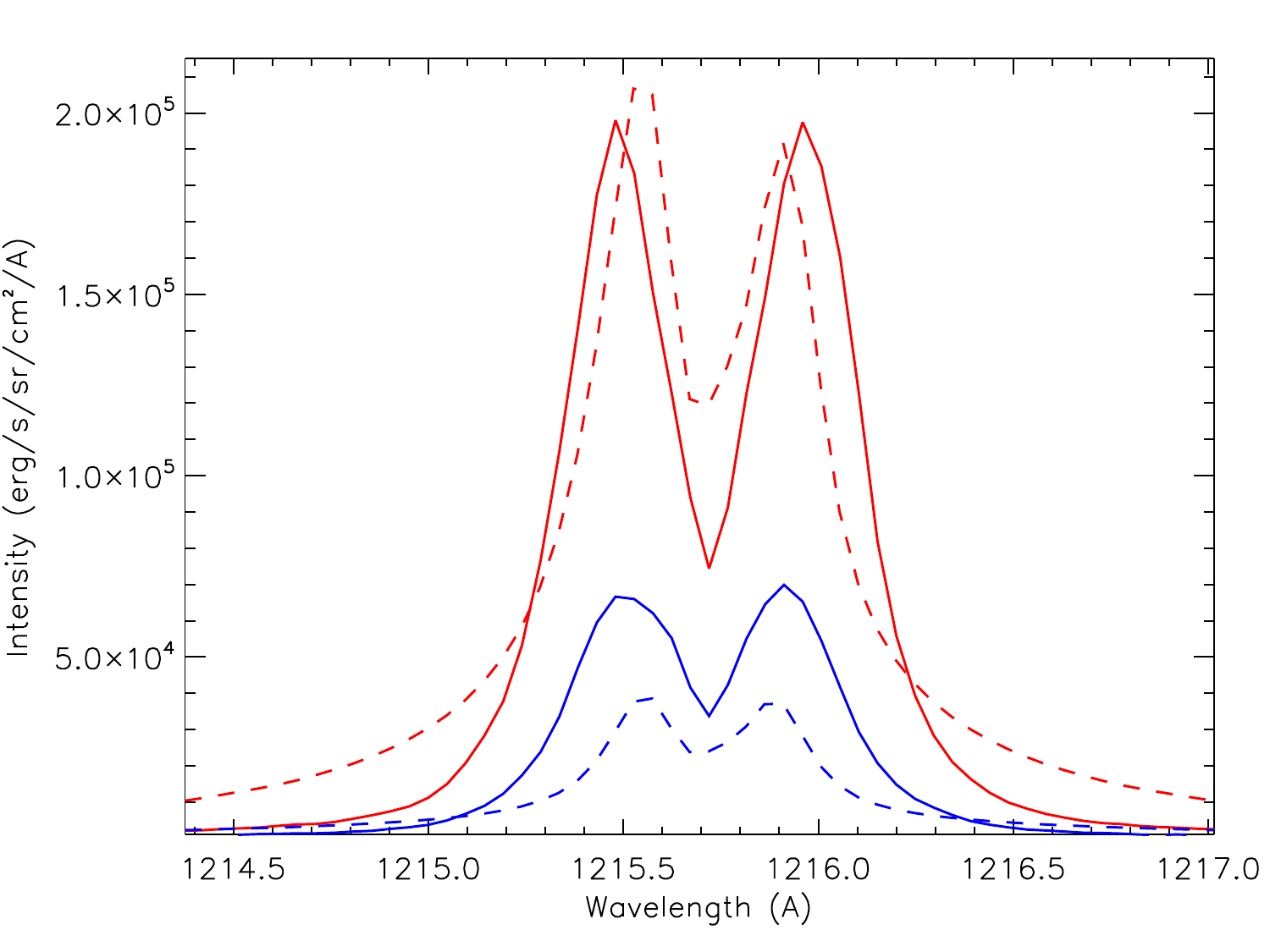}
\caption{ CLASP profiles near the network, Y=46" (red solid), and in the internetwork, Y=33" (blue solid) at t=25 s. Synthetic Ly-$\alpha$ profiles from network (red dashed) and internetwork (blue dashed). The location of the Bifrost profiles are labeled in Figures 1 and 9.}
\label{fig:mpr}
\end{figure}
\citet{golding_16}.
The magnetic field in the simulation has a bipolar configuration with an unsigned field strength of 50G in the photosphere. The simulation setup is otherwise identical to the one in
\citet{carlsson_16}
and we refer to that paper for details.
We select a single snapshot from the simulation at $t=1000$~s that we use as input atmosphere for the subsequent radiative transfer computations. To save computation time we downsampled the atmosphere to a grid of $252 \times 252 \times 496$ points.
 The emergent spectra are calculated in full 3D non-LTE with the Multi3d code 
\citep{leenaarts_09b}. 
The \HI\ and \MgII\ model atoms are the same as described in 
\citep{sukhorukov_17}, 
and include partial frequency redistribution for the Mg II h\&k lines as well as Ly-$\alpha$ and Ly-$\beta$.
This treatment of the lines is essential for a physically accurate modeling of strong resonance lines formed in the chromosphere and transition region.
The emergent radiation that we use in our analysis in Section~\ref{sec:analysis} was stored for rays parallel to the $x$-axis and parallel to the $y$-axis for all vertical inclinations $\mu = \cos\theta$ between 0.2 and 1.0 in steps of $\Delta \mu=0.1$.\\
\indent We show the photospheric magnetic field configuration as well as example images in Fig.~\ref{fig:sim}. The magnetic field in the photosphere shows a large-scale bipolar configuration, with the field concentrated in the intergranular lanes. The Ly-$\alpha$ and \MgII~h panels are dominated by fibrils that emanate from the photospheric field concentrations.  The lower right panel shows the frequency-integrated Ly-$\alpha$ intensity, which can be compared with Fig.~8  of
\citet{2010SoPh..261...53V},
which shows a line-integrated  Ly-$\alpha$ image of a supergranular cell interior obtained with the VAULT rocket experiment.
%A detailed description of the simulation and the synthesized radiation is being written (Sukhorukov et al., in prep.).
%%%%%%%%%%%%%%%%%%%
\section{Analysis} \label{sec:analysis}
%%%%%%%%%%%%%%%%%%%
The full field of view of the CLASP SJ is shown in Figure  \ref{fig:fov}, along with a simultaneous Solar Dynamics Observatory/Helioseismic and Magnetic Imager magnetogram \citep{scherrer_12}.
The CLASP slit can be seen at image center.
\begin{figure}
\includegraphics[width=0.5\textwidth]{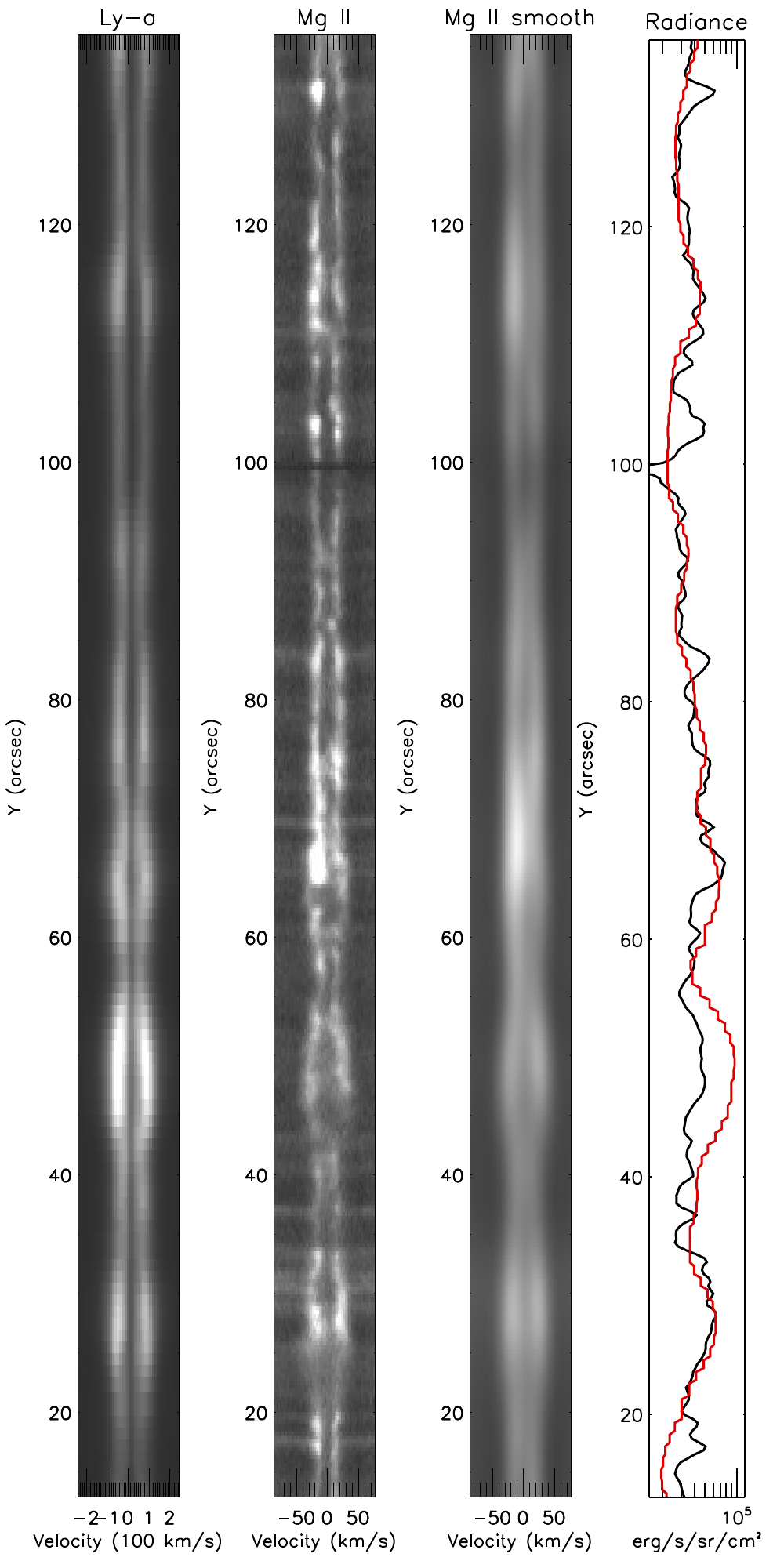}
\caption{Example spectra along the slit. CLASP Ly-$\alpha$ (left image), IRIS Mg II h (middle image), Mg II h smoothed to 2.8" and 26 km/s (right image). The CLASP exposure is from t=103 s. The line plot shows the integrated line radiance for Ly-$\alpha$ (red, $\pm 0.96$ \AA ) and Mg II h (black, $\pm 0.75$ \AA, the Mg II radiance has been reduced by a factor of 4 for ease of display)}
\label{fig:exspec}
\end{figure}
The CLASP slit encountered almost exclusively internetwork or weak network magnetic concentrations.
The only exception is a strong positive flux concentration near Y=40".
The Ly-$\alpha$ profiles are quite variable and two examples are shown in Figure \ref{fig:mpr}.
Ly-$\alpha$ is an optically thick transition.
The observed intensity for each spatial and spectral position is tied to the radiation field and plasma distribution along the line of sight which determine the altitude of the last scattering surface.
The characteristic shape, double peaked with a depressed core, is common among strong chromospheric/transition region emission lines.
This profile shape is generated by the non-LTE formation of the line, where the line source function is mainly determined by scattered radiation and  decouples from the increasing (as a function of altitude) Planck function in the chromosphere \citep{avrett_08}.\\
\indent The largest dataset of solar Ly-$\alpha$ profiles was collected by the LPSP instrument aboard OSO-8 \citep{bonnet_78,gout_78}.
An averaged LPSP profile is shown in blue in Figure \ref{fig:ex_ly} (spatial bin of 1"$\times$10").
\begin{figure*}
\includegraphics[width=\textwidth,trim=1cm 5.2cm 1cm 5cm,clip]{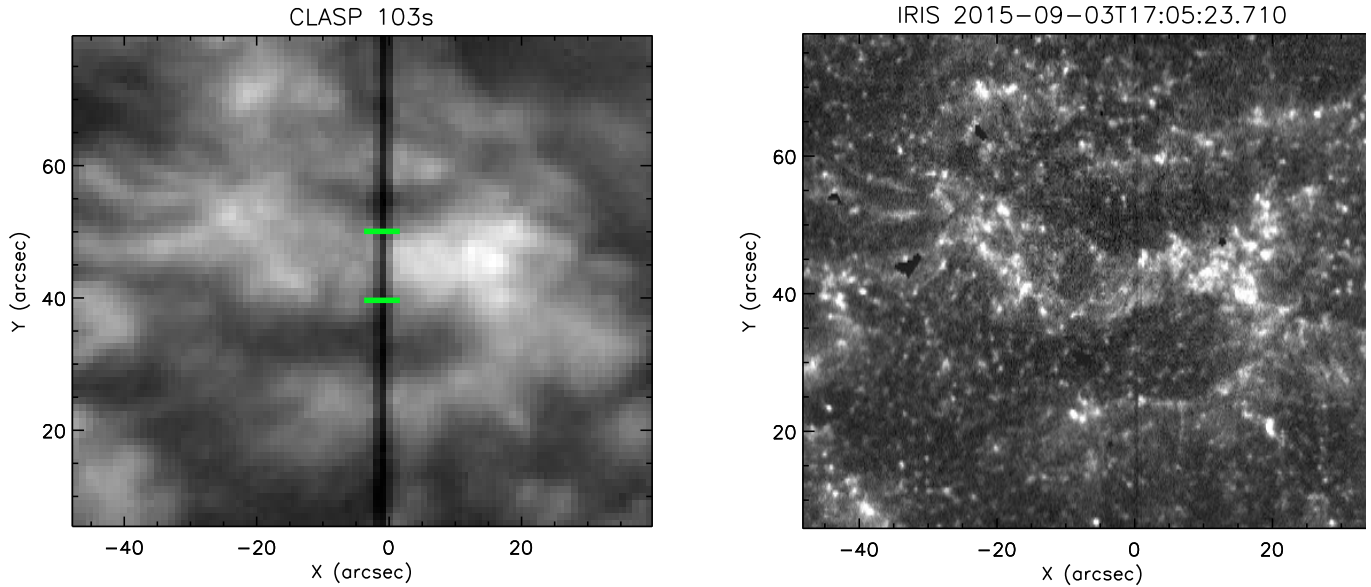}
\caption{A strong magnetic region covered by both IRIS and CLASP slits. The CLASP slit jaw bandpass is centered on 1215\AA~while the IRIS bandpass is centered on 1400\AA. The overlapping spectrograph data is shown in Figure~\ref{fig:fits}. The green bars show the position of the spicule features shown highlighted in Figure~\ref{fig:fits}. An animated version of this figure is available in the electronic version of the paper.}
\label{fig:sj}
\end{figure*}
%The profile shape in the wing and the separation (i.e. width) between the L2 peaks are agree well between CLASP and LPSP.
LPSP was able to resolve the geocoronal absorption line.
CLASP cannot so the intensity at the L3 is depressed relative to the radiated intensity.
Given the absorption profiles observed by LPSP (albeit at 500 km altitude and not 250 km like CLASP) and CLASP's spectral resolution, the CLASP cores are consistent with LPSP.\\
%The asymmetry statistic is defined as:
 %\begin{displaymath}
%A_L=(I_{L2V}-I_{L2R})/(I_{L2V}+I_{L2R}).
%\end{displaymath}
%Both the LPSP data and the CLASP have mean asymmetries of 2.5\%.
%The time-dependent absorption factor does a net effect on asymmetry which will be described in Section 3.2.
%Asymmetry is a particularly variable statistic with regards to solar structures so the it is sensitive to the averaging field of view.\\
\indent Figure \ref{fig:exspec} shows the co-spatial, co-temporal profiles of both Ly-$\alpha$ and Mg II h.
In this analyze, we focus on Mg II h and not k because the line fitting technique described in \cite{schmit_15} is affected by the Mn I 2795.6\AA~line.
The only previous simultaneous measurement of both lines is detailed in \cite{bocch_94}.
That analysis was conducted primarily using 10" resolution data.
%Our dataset is both more comprehensive and higher resolution.
When compared side-by-side, the Ly-$\alpha$ and Mg II h spectra have many similarities.
%The line are predicted to form in relatively close vertical proximity to each other.
%The VAL atmosphere \citep{val_81} suggest that the L3 and h3 cores form within 300 km.
The two easiest parameters to compare by eye are the peak intensities ($I_{L2}, I_{h2}$) and peak widths (i.e. $\lambda_{L2r}-\lambda_{L2v}$).
The broadest profiles tend to occur cospatially for both lines.
\begin{figure*}
\centering
\includegraphics[width=.85\textwidth]{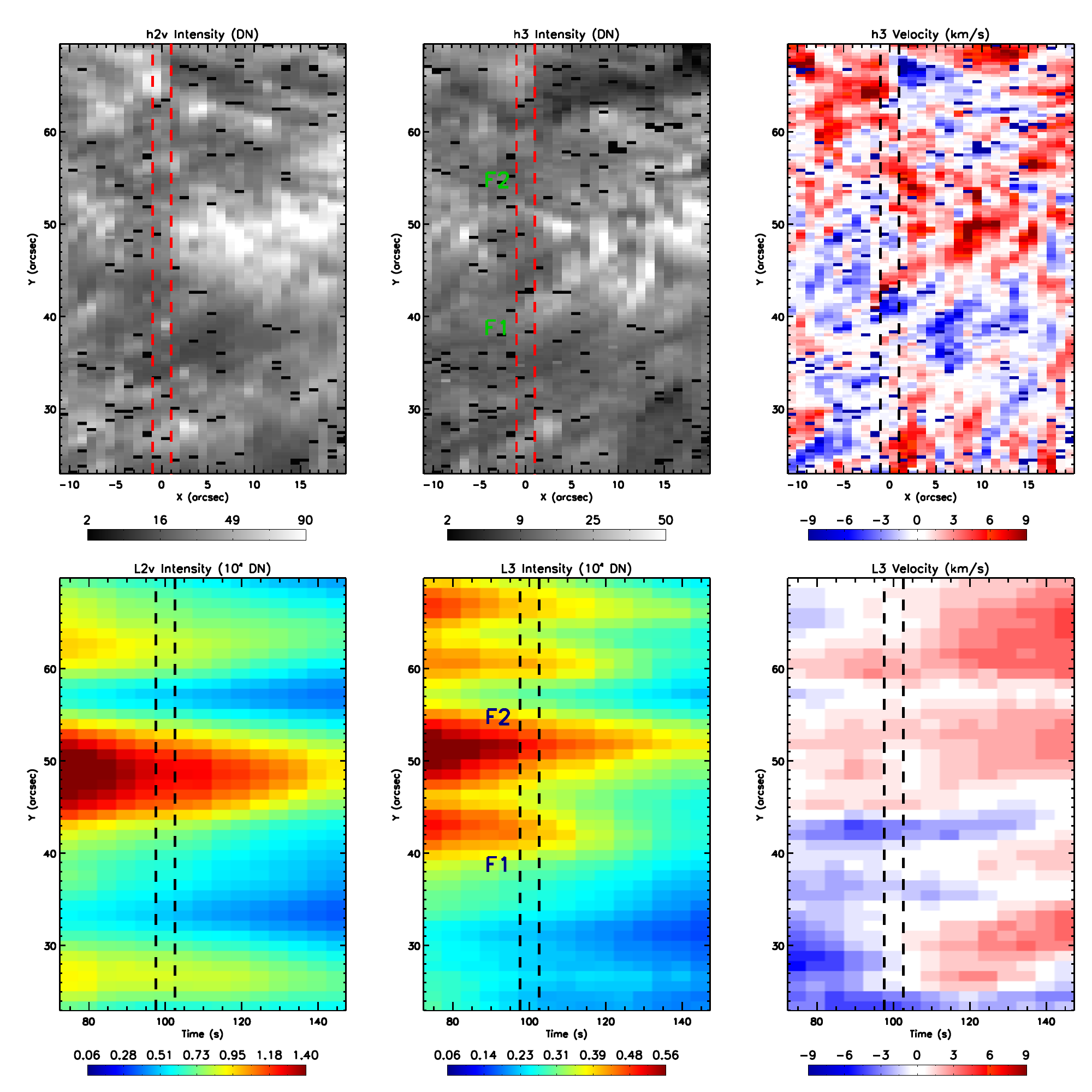}
\caption{Comparing the co-spatial and co-temporal profiles in Mg II h and Ly$\alpha$. The red/black dotted lines mark the spatial/temporal region where the rastering IRIS slit overlaps the stationary CLASP slit. Violet peak intensity (left column), core intensity (middle column), and core velocity (right column). Note that the Mg II data is a raster scan, with both spatial position and time changing along the $x$-axis, while the Ly-$\alpha$ data are sit-and-stare, and the $x$-axis shows time variation at a fixed slit location. The time-dependent absorption affects both the intensity and velocity of the Ly-$\alpha$ fits. The effect is approximately uniform along the slit.}
\label{fig:fits}
\end{figure*}
With regards to peak intensities, we find that there is significantly more complex structure in the Mg II profiles in both the spatial and spectral dimensions.
Part of this effect is related to the difference in resolution between the CLASP and IRIS datasets.
To get a sense for the resolving power effect, we have convolved the Mg II spectra with a two-dimensional gaussian (full width at half maximum, FWHM, spectral:  26 km/s, FWHM spatial: 2.8").
At the resolution of CLASP, the great complexity of the Mg II spectra is reduced although profile asymmetries are still apparent, as are variations in peak width and integrated intensity.
%The convolution does not lead to a perfect match of the data.
%We find that strong asymmetries in Mg II are more subdued in Ly-$\alpha$.
%This is visible in Figure \ref{fig:exspec} at Y=68" and Y=103".
%Asymmetric Mg II profiles are associated with chromospheric plasma motions \citep{leenaarts_13b}.
%This comparison may be revealing that some chromospheric velocity perturbations do not reach the transition region or that the transition region along this line of sight is disjoint from chromospheric region visible in Mg II. 
There are also small offsets between the spatial positions of bright $L2$ profiles compared to the bright $h2$ counterparts.
The feature between $43"<Y<55"$ is an example where the emission enhancement Ly-$\alpha$ is more extended than in the Mg II lines.
%If we consider emission along spicules or loops inclined to the vertical, tfdhe Ly-$\alpha$ emission is either more extended or weighted 
\subsection{Magnetic Region}
Figure \ref{fig:sj} shows a subregion of the CLASP and IRIS slit jaw data at higher resolution.
An animation is available in the electronic version of this article.
This magnetic region is the largest one crossed by the CLASP slit.
%The CLASP slit jaw imager has spatial resolution of 2" while the asymmetrically binned IRIS slit jaw data has resolution of  0.16"x0.33" in the X and Y directions respectively.
The CLASP slit jaw imager has spatial resolution of 2" while the IRIS slit jaw data has resolution a spatial resolution of 0.4".
The IRIS 1400 \AA~bandpass image contains a mixture of continuum and line emission \citep{jms_15b} so bright structures are a combination of extended transition region structures (fibrils and spicules and high altitude shocks) and temperature-minimum/low-chromospheric structures (low altitude shocks and vertically-oriented magnetic flux tubes).
The clearest elongated structures extend in the negative-X direction from $X$=-24" for $45"<Y<65"$.
As measured by \cite{pereira_14}, the lifetime of spicules is generally around 150-200s.
Our short duration movie does not capture any dramatic dynamics, but the movie shows ubiquitous flows and localized brightenings over and surrounding the magnetic region.
Over the section of the magnetic concentration covered by the slit, there appears to be some magnetic restructuring near $(X,Y)$=(4,45) that reaches quiescence by $t$=117s.
We highlight this region in Figure \ref{fig:fits}.\\
\indent We have plotted three congruent statistics for Ly-$\alpha$ and Mg II h overlying the magnetic region.
However the two datasets are not identical.
As described in Section 2, IRIS is undergoing a spatial raster while CLASP maintained a constant pointing.
The red dashed lines (top panels) and black dashed lines (bottom panels) represent the spatial region and time period where the CLASP and IRIS slits overlapped.
The CLASP profile fits are affected by the time-dependent absorption.
Intensity decreases as a function of time, and velocities are increasingly red-shifted.
The absorption effect is approximately uniform along the slit.\\
\indent The two most interesting structures in the FOV are labeled F1 and F2.
Based on the slit jaw images and the $I_{h3}$ spectroheliogram, we suggest that these structures are fibrils or spicules.
These are complicated structures.
They are bright in the slit jaw and $I_{h3}$ and $I_{L3}$.
They are dim at $I_{L2v}$ and $I_{h2v}$.
F1 exhibits a strong blue shift in both lines but at F2 the core velocity is zero.
The magnetic field in this region is concentrated into one large element which is surrounded by internetwork (at the resolution limit of HMI).
We expect fibrils and spicules to extend outward from this magnetic concentration.
F1 and F2 are likely rooted in the strong magnetic region and exhibit enhanced heating in the transition region and chromosphere.
In scanning from line core toward the line wing, we observe deeper into the atmosphere.
At $\lambda_{h2v}$ and $\lambda_{L2v}$, we are likely seeing regions formed in the internetwork which are cooler than the magnetic concentration.
We do not see Doppler shifts at L3  higher than 10 km/s as have been reported in spicules or rapid blueward excursions \citep{rouppe_09}.
\cite{sekse_13} found that waves and not just bulk flows contribute to the rapid blueward excursions.
For bulk flows, structures that are inclined relative to the line of sight may only have a fractional velocity component projected along the line of sight.
F1 and F2 are at a viewing angle $\mu=0.6$. 
The chromospheric velocity field is expected to be highly structured on small scales ($\leq$ 3" similar to granulation) over characteristic lifetimes of $\geq 100$ seconds.
Spicules can only be clearly resolved at spatial resolutions $<0.5"$ so our middling Doppler shifts may just be attributed to spatial smoothing.
\begin{figure*}
\centering
\includegraphics[width=\textwidth,trim=2cm 16.5cm 2cm 2cm,clip]{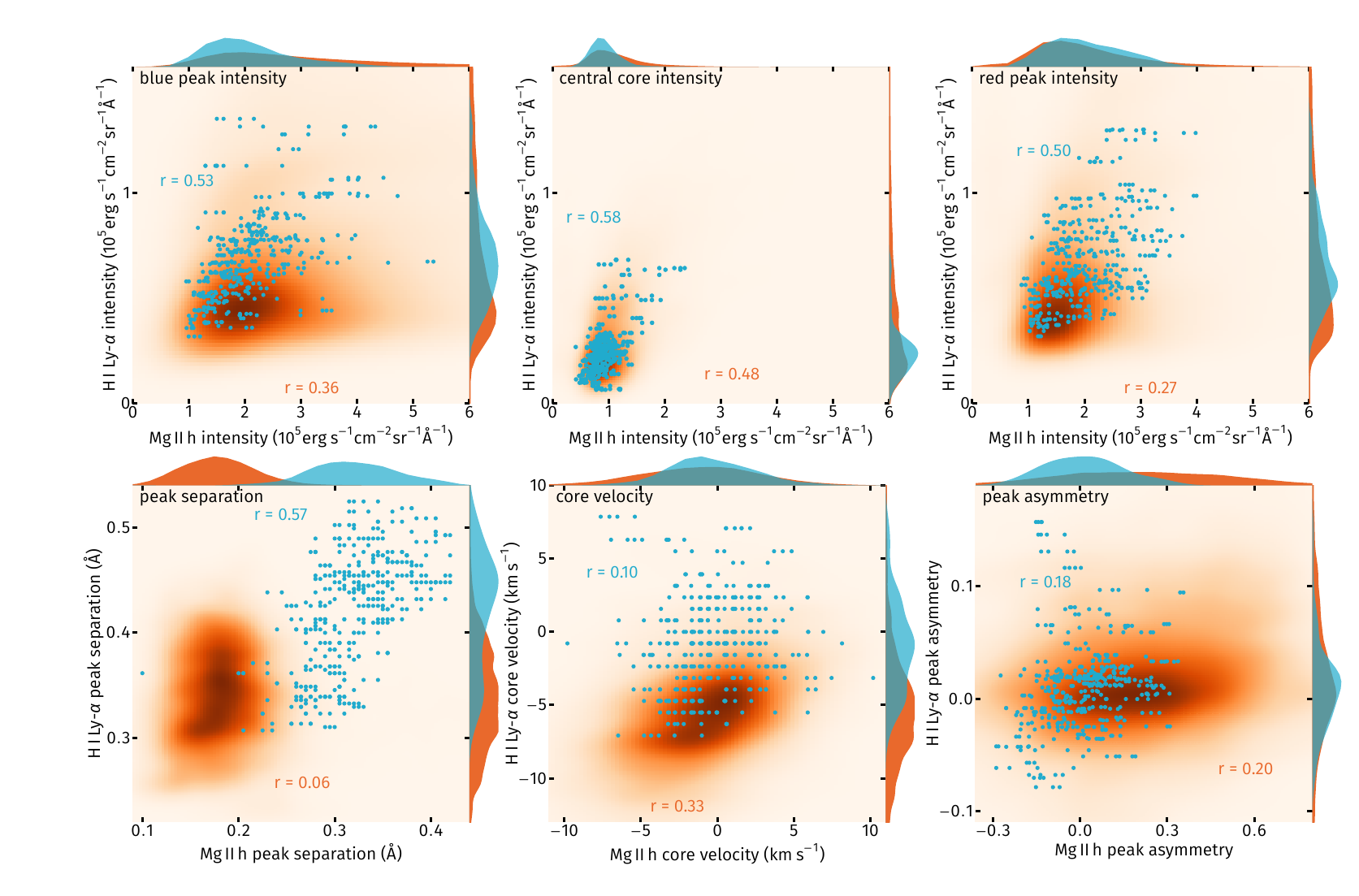}
\caption{Examining congruent statistics between co-spatial, co-temporal Mg II h and Ly-$\alpha$ profiles. The teal dots are taken from the observations. The background contours are the joint probability distribution computed from the Bifrost model atmosphere. Pearson coefficients for the data and the simulation are shown in teal and orange respectively. The teal and orange histograms represent the observed and simulated distributions, respectively, of the statistic shown on that axis.}
\label{fig:scat1}
\end{figure*}
%Given the very different resolution of the CLASP and IRIS instrument, it is difficult to determine whether on smooth structure in Ly-$\alpha$ is reflective of difference between the chromosphere and the transition region or simply an instrumental effect.
%Wave theory suggests that many of the oscillations that affect the chromosphere are reflected in or below the transition region \citep{erdelyi_07} which might result is a more smoothly structured H I profiles.
\subsection{Statistical Comparison of Profiles}
Figure~\ref{fig:scat1} presents scatter plots displaying the correlation between the profile parameters for the Mg II and Ly-$\alpha$ lines.
We use 120 CLASP profiles (drawn from along the slit) and (the overlapping) 487 IRIS profiles, that span view angles $0.33<\mu<0.62$.
These profiles are from a single co-added CLASP exposure (t=25s) so the instrumental absorption feature, while it is present, will not biasedly scatter the intensity distributions.
The scatter points overlie the 2D joint probability distribution of the parameters extracted from the synthesized Bifrost spectra.
A geocoronal absorption line is added to the synthetic spectra as a Gaussian line with a relative depth of 60\% and FWHM = 27 m\AA, centered at the nominal Ly-$\alpha$ central wavelength and corresponding to a thermal absorption of an optically-thin slab of hydrogen with $T = 10^3$ K \citep{bruner_69}.
The synthetic spectra were smoothed and then binned to match the resolution and plate scale of the two instruments.
The synthetic spectra were fit using the same technique as the data.\\
\indent The correlations between $I_\mathrm{h2v}$ and $I_\mathrm{L2v}$, and  $I_\mathrm{h2r}$ and $I_\mathrm{L2r}$ are similar.
We have relatively high coefficients (Pearson $r$-value) for the observed lines and lower values in the synthetic lines.
%The synthetic emission in Ly-$\alpha$ is dimmer than the estimated observed emission.
%The CLASP data has been \jledit{calibrated using the OSO-8 data, but given the water vapor contamination we do not put much weight behind the absolute intensity calibration (see Section 2.1). If the discrepancy between the observed and modeled peak intensities is real, then this points to deficiencies in the modeling. Possible reasons for this could be that: 1) the Bifrost model is too cold in the upper chromosphere and transition region. 2) The Ly-$\alpha$ intensities should be computed using full non-equilibrium radiative transfer instead of the in statistical equilibrium as we have done here. \citet{golding_17} showed that this is required for the \ion{He}{2} 256~\AA\ and 304~\AA\ lines. 3) Particle diffusion of neutral hydrogen should be included \citep{1990ApJ...355..700F}.}
% 
In the synthetic data, there is a broader spread in Mg II h2v intensities than h2r, but this trend is not duplicated in Ly-$\alpha$.
The positive peak asymmetry (bright 2v and dim 2r) is not as pronounced in Ly-$\alpha$ as Mg II h.\\
%This is an indication of the skewness of the vertical velocity distribution in the chromosphere \citep{leenaarts_13b}.
%The shocks that are responsible for much of the asymmetry in Mg II profiles spe \\
%Differences in $I_{h2v}$ and $I_{h2r}$ have been related to velocity fields in the upper chromosphere.
%The variability of peak emission is linked with the distribution of vertical velocities and the correlation between velocity and column mass in the upper chromosphere.\\
%We find there is a stronger correlation between the intensity at $h3$ and $L2$ than at $h2$ and $L3$.
%This is logical given that Mg II should be formed lower than Ly-$\alpha$, and that line core emission is formed higher than peak emission.
%We see no correlation in the base width, although this is not surprising given our narrow spectral window for Ly-$\alpha$.
\indent Our observed profiles are about a factor of two broader than the synthetic ones.
This is a documented characteristic of the current generation of Bifrost models \citep{leenaarts_13a}.
While we find a high degree of correlation in the peak width of the observed lines, the synthetic lines have no correlation.
Synthetic Ly-$\alpha$ peak width is more variable than synthetic Mg II peak width.
As described in \cite{leenaarts_13b}, the peak width is an indication of non-monotonic vertical velocity near the line centroid $\tau=1$ layer.
The profile width for both lines is also affected by the vertical profile of the source function.
If the maximum of the source function is pushed to lower altitudes, the peak width of the line will increase.
The lack of correlation in the simulation points to physical circumstances acting in the chromosphere that are not properly present in the simulations, and thus provides an additional test of the models.
To which extent the lack of correlation has to do with differences in formation height, the velocity field, turbulence and/or the source function (and thus ultimately temperature) requires further investigation.\\
%More generally, line profile width is tied to the growth in optical depth below the line core $\tau=1$ layer.
%If we consider Gaussian emission/absorption profiles as a probability function, it is unlikely that photons are created/destroyed in the wings.
%When the core of the line is extremely optically thick, these wing photons that are created infrequently are still able to escape unscattered from the column at a much higher rate than line core photons.
%This intuitive explanation is sometimes called ``opacity broadening''.
%Physical processes that add cool plasma to the upper atmosphere could produce additional broadening.
%The low degree of correlation in the peak width statistic for the synthetic lines implies that either the velocity fields or the source function profiles are quite different in the mutual formation range of the two lines.\\
%This correlation implies that the spectrum of unresolved flows at the upper chromosphere and the upper transition region are not well correlated in the model.\\
%We do not expect that the geocoronal feature to effect the observed Ly-$\alpha$ peak width.
%The low $r$-value for the synthetic peak width is an indication that the velocities fields in the transition region and chromosphere are distinct and in many instanced decoupled.\\
%To further study that relationship, models that contain both bulk flows and Doppler oscillations are needed.
\indent In the observed data, we find a weak correlation in core velocity of the lines (r=0.1).
That may partially be explained by spectral resolution as the typical velocities are under-resolved.
In the model, we do see a better correlation between the velocities derived from the profiles.
Although the synthetic spectra are degraded to the instrumental resolution, they do not include noise which might limit the accuracy of the velocity determination.
%The core velocity is expected to map well with the vertical velocity at the $tau=$1 layer.
%Given the different formation heights of the lines, it is reasonable to expect that simultaneous spectra will not display identical signatures of vertically propagating disturbances.
%The Ly-$\alpha$ profile tend to be more symmetric than the Mg II profiles, although the average profile of for both lines is violet-side dominant.
%In fact there is a total dearth of strongly red-side dominant profiles in the Ly-$\alpha$ dataset, where as in the Mg II profiles they make up approximately 8\% of the ensemble.
\subsection{Velocity Field in Bifrost}
\begin{figure*}
\centering
\includegraphics[width=.95\textwidth,trim=2.5cm 12cm 2.5cm 10cm,clip]{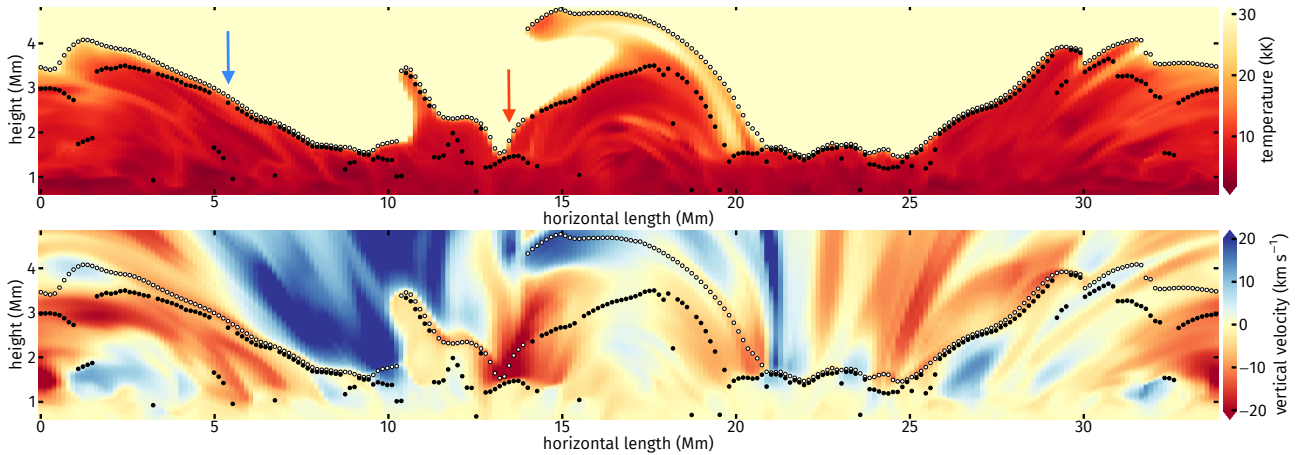}
\caption{A vertical slice through the Bifrost atmosphere showing temperature (top panel) and vertical velocity (bottom panel). The white circles mark the altitude of Ly-$\alpha$ L3 $\tau=1$ layer while the black circles show the same for Mg II h3. The red and blue arrows mark profiles shown in Figure 4. The position of slice in the Bifrost box is displayed by the yellow line in Figure 1.}
\label{fig:slice}
\end{figure*}
The formation of Mg II h and Ly-$\alpha$ in the solar atmosphere were predicted to be similar based on canonical models \citep{val_81,avrett_08}.
In such hydrostatic models, Mg II h3 is expected to map to the top of the chromosphere where the gas temperature is $1.0-2.0\times10^4$ K.
Ly-$\alpha$ L3 is expected to map to the transition region where the gas temperature is $2.0-4.0\times10^4$ K.
Ly-$\alpha$ almost by definition demarcates the edge of the transition region because once hydrogen is fully ionized the temperature gradient is expected to dramatically rise until a coronal equilibrium is reached.
Both Mg II h and Ly-$\alpha$ are resonance transitions so the profile wings are expected to map continuously throughout the chromosphere to the altitude where the background continua are formed.\\
\indent Based on those predictions, our hypothesis upon beginning this analysis was that the Mg II and H I profile parameters would be highly correlated.
The observations do not match that hypothesis, particularly for the statistics tied to velocities.
The Bifrost model allows us to examine a simulated solar-like atmosphere and investigate how the plasma properties in that atmosphere combine to create synthetic emergent line profiles.
Figure~\ref{fig:slice} shows a vertical slice of from the Bifrost atmosphere.
The $\tau=1$ altitude for L3 at $\mu=1$ is marked by white circles while the $\tau=1$ altitude for h3 is marked by black circles. The slice has photospheric magnetic concentrations between  $8<x<13$~Mm and $21<x<24$~Mm comparable to solar network elements, while the rest of the atmosphere contains loops and quiet areas more reminiscent of internetwork quiet Sun.
In many areas the formation heights of the central depression of both lines are very close together, in line with the predictions of semi-empirical 1D models \citep{val_81,avrett_08}. 
This is in particular the case in the range $21<x<24$~Mm. Here the transition from chromospheric temperatures is very steep, and the chromosphere has a relatively high mass density (not shown), leading to the $\tau=1$ heights for both lines to be located closely together.
However, there are also many areas where the formation heights lie further apart, such as $0<x<3$~Mm and $14<x<20$~Mm. The vertical velocities at the formation heights of the lines in those areas are typically very different. The reason for the formation height difference are the atomic structure and abundance differences between Mg II and H I. Hydrogen is roughly a factor of $2\times10^4$ more abundant and neutral hydrogen does not ionize to higher ionization states as readily as Mg II \citep[see for example][]{carlsson_12,rutten_16, rutten_17}.
The chromosphere and lower transition region in the internetwork regions of the simulation is extended, has a low mass density and temperatures ranging from $1\times10^4$ K to $3\times 10^4$ K. In such regions Ly-$\alpha$ will reach optical depth unity high in the atmosphere, but for the Mg II~h line optical depth unity will lie much deeper. We speculate that the lack of correlation between the observed L3 and h3 velocities is caused by this effect because the observations targeted predominantly internetwork.
In comparing the simulation's velocity field with that of the Sun, it is important to note that the simulation does not produce spicules.
Spicules are extensions of 1-2$\times10^4$ K plasma, a few megameter long, that move 20-60 km/s upwards from the chromosphere.
This may be tied to resolution or neglected terms in the equations being solved.
Spicules may change the mass distribution in the simulation atmosphere as well as the velocity field.
%We can see that thermal structures are related to magnetic structures, while the vertical velocity structure is significantly more complex.
%Between 2-4 Mm in altitude, the thermal structure is bifurcated: regions where $T<1\times10^4$ K and regions where $T>2\times 10^4$ K. 
%The boundary between these regions is largely defined by horizontal magnetic loops.
%The formation heights of h3 and L3 are coupled to density and temperature more closely than velocity.
%We can see that the difference in formation height is different in different magnetic structures, and that the velocities at these surfaces are often not equal.
\section{Conclusions}
Our Mg II/Ly-$\alpha$ observations are the highest resolution joint dataset to date.
They are also the first dataset collected with high resolution magnetograms and chromospheric images to provide context. 
For these reasons, these data are an important window into the connection between the chromosphere and the transition region.
Our analysis of these data is affected by the limitations of the observations.
As a rocket-borne instrument, CLASP can only collect data for minutes.
To boost the polarimetric signal, the instrument's spectral resolution is coarse and the pointing is constant.
In addition to the unavoidable (outside heliocentric orbit) geocoronal feature, the CLASP data also contains a broad molecular absorption feature.
To achieve co-pointing with CLASP, IRIS observed in a wide-field rastering mode.
Therefore the slit alignment cadence is non-ideal to conduct a study of dynamics.
We have used our joint dataset to analyze the regional and statistical correlations between the spectral profiles.
When we smooth IRIS data to CLASP's spatial and spectral resolution, the profiles look qualitatively similar.
There is likely a great deal of structure in Ly-$\alpha$ profiles, only visible at higher resolution, that might provide useful diagnostics for future observatories.
While there are general similarities between the lines there are many differences as well.
One explanation for the variations can be tied to variation in formation height that is predicted by Bifrost.
While the temperature difference of the h3 and L3 formation layers may only be 1$\times 10^4$ K, the magnetic field lines or velocity streamlines that thread those layers may be completely unrelated.\\
\indent It is not wholly unexpected that the simulation and data do not perfectly agree on how Mg II statistics are related to those for Ly-$\alpha$.
The number of profiles we have in our sample is too small for a proper statistical study.
We are not able to probe temporal variations, and we know what the chromosphere and transition region are dynamic over timescale of $<10$s \citep{kubo_16}.
We know that the Bifrost model atmosphere does not reproduce all characteristics of other chromospheric and transition region diagnostics.
The line profiles of the Mg II h and C II (1334.5\AA~and 1335.6\AA) are too narrow \citep{leenaarts_13b,rathore_15} and the transition region profiles are too dim \citep{schmit_16}.
The simulation lacks the signatures of spicules.
There are indications that a more complete treatment ion-neutral-electron interactions eliminates some of these discrepancies \citep{jms_16}.
These are the limitations that we are presented with.\\
%Our statistical sample is not large enough to allow us to further break down our datasets with more refined groups.
%We have examples of spectra overlying one large magnetic regions, but a majority of the slit is only incident on weak network concentrations.
%In our magnetic region, the slit jaw data show fibrils and spicules in both Si IV and Ly-$\alpha$ passbands.\\
\indent Ly-$\alpha$ is an important spectral line for understanding the dynamics and energy balance at the base of the corona.
Rocket payloads are often a testbed for technological demonstrations.
Enhanced spectral resolution, a large dynamic range, and broad spectral window to sample the photosphere/temperature minimum would be ideal instrument capabilities to strive for.
In preparation for these measurements, we need to explore the linkage between the chromosphere and the transition region using the advanced radiative MHD models which are able to capture a broad range of observed phenomena.
\section*{}
We acknowledge the Chromospheric Lyman-Alpha Spectropolarimeter (CLASP) team. The team was an international partnership between NASA Marshall Space Flight	 Center, National Astronomical Observatory of Japan (NAOJ), Japan Aerospace	 Exploration Agency (JAXA), Instituto de Astrofísica de Canarias (IAC) and Institut d'Astrophysique Spatiale ; additional partners include Astronomical Institute ASCR,	 Lockheed Martin and University of Oslo. US participation is funded by NASA Low Cost Access to Space (Award Number 12-SHP 12/2-0283). Japanese	participation is funded by the basic research program of ISAS/JAXA, internal research	funding of NAOJ, and JSPS KAKENHI Grant Numbers	23340052, 24740134, 24340040, and 25220703. Spanish participation is funded by	the Ministry of Economy and	Competitiveness through project AYA2010-18029 (Solar Magnetism and Astrophysical Spectropolarimetry).	French	hardware participation was	funded by Centre National d'Etudes	Spatiales (CNES).\\
\indent IRIS is a NASA small explorer mission developed and operated by LMSAL with mission operations executed at NASA Ames Research center and major contributions to downlink communications funded by ESA and the Norwegian Space Centre.\\
Computations presented in this paper were performed on resources provided by the Swedish National Infrastructure for Computing at the National Supercomputer Centre at Link\"oping University, at the PDC Centre for High Performance Computing at the Royal Institute of Technology in Stockholm,  and at the High Performance Computing Center North at Ume\aa\ University.

%\begin{table*}
%\begin{tabular}{c | c}
%Variables & $r$\\
%\hline
%\hline
%$\bar{I}_{h2}$ v. $\bar{I}_{L2} $& 0.56\\
%$I_{h3}$ v. $I_{L3}$ & 0.58\\
%$\bar{I}_{h2}$ v. $I_{L3} $& 0.26\\ 
%$I_{h3}$ v. $\bar{I}_{L2} $& 0.46\\
%\hline
%\hline
%Wid h2 v. Wid L2 & 0.57\\
%$\int I_{Mg}$ v.  Wid L2 & 0.24\\
%$I_{h3}$ v. Wid L2 & -0.28\\
%\hline
%\hline
%$\int I_{Mg}$ v. $\int I_{Ly}$ & 0.59\\
%\end{tabular}
%\caption{\label{tab:corr} List of interesting correlation coefficients. $\bar{I}$ is the average of the v- and r-peaks.``Wid" denotes the peak width. $\int{I}$ is the integrated radiance.}
%\end{table*}

\end{document}